\documentclass[pre,twocolumn,aps,superscriptaddress]{revtex4}
\usepackage[T1]{fontenc}
\usepackage[latin9]{inputenc}
\usepackage{array}
\usepackage{booktabs}
\usepackage{mathrsfs}
\usepackage{mathbbol}
\usepackage{multirow}
\usepackage{amsmath}
\usepackage{amsthm}
\usepackage{amssymb}
\usepackage{stmaryrd}
\usepackage{graphicx}
\usepackage{latexsym}
\usepackage{textcomp}
\usepackage{mathtools}
\usepackage{xcolor}

\usepackage[T1]{fontenc}
\usepackage{adjustbox}
\usepackage{scalerel}
\usepackage{braket}

\usepackage[citecolor=magenta,colorlinks=True]{hyperref}
\usepackage{lineno}
\usepackage{stackrel}
\usepackage[toc,page,titletoc]{appendix}
\usepackage{bm}
\definecolor{mycolor1}{rgb}{0.1, 0.6, 0.6}

\begin{document}

\title{Instabilities govern the low-frequency vibrational spectrum of amorphous solids}

\author{Surajit Chakraborty} 
\email{schakraborty@tifrh.res.in}

\author{Roshan Maharana} 
\email{roshanm@tifrh.res.in}

\author{Smarajit Karmakar} 
\email{smarajit@tifrh.res.in}

\author{Kabir Ramola} 
\email{kramola@tifrh.res.in}

\affiliation{Tata Institute of Fundamental Research, Hyderabad 500046, India}


\begin{abstract}
Amorphous solids exhibit an excess of low-frequency vibrational modes beyond the Debye prediction, contributing to their anomalous mechanical and thermal properties. Although a $\omega^4$ power-law scaling is often proposed for the distribution of these modes, the precise exponent remains a subject of debate. In this study, we demonstrate that boundary-condition-induced instabilities play a key role in this variability. We identify two distinct types of elastic branches that differ in the nature of their energy landscape: Fictitious branches, where shear minima cannot be reached through elastic deformation alone and require plastic instabilities, and True branches, where elastic deformation can access these minima. Configurations on Fictitious branches show a vibrational density of states (VDoS) scaling as $D(\omega) \sim \omega^3$, while those on True elastic branches under simple and pure shear deformations exhibit a scaling of $D(\omega) \sim \omega^{5.5}$. Ensemble averaging over both types of branches results in a VDoS scaling of $D(\omega) \sim \omega^4$. Additionally, solids relaxed to their shear minima, with no residual shear stress, display a steeper scaling of $D(\omega) \sim \omega^{6.5}$ in both two and three dimensions. We propose two limiting behaviors for amorphous solids: if the system size is increased without addressing instabilities, the low-frequency VDoS scales with an exponent close to $3$. Conversely, by removing residual shear stress before considering large system sizes, the VDoS scales as $D(\omega) \sim \omega^{6.5}$.
\end{abstract}

\maketitle

{\large\textbf{Introduction}}

Low-frequency vibrations in solids influence various mechanical and thermal properties, including their low-temperature specific heat and thermal conductivity~\cite{zeller1971thermal,alexander1998amorphous,pohl2002low,ramos2022low}. In crystalline structures, vibrational modes (phonons) follow a low-frequency vibrational density of states (VDoS) described by the Debye model as $D(\omega) \sim \omega^{d-1}$ in $d$ dimensions, which explains the observed specific heat and thermal conductivity in crystals~\cite{Debye_1912,kittel2005introduction}. Conversely, amorphous solids exhibit anomalous mechanical and thermal properties compared to their crystalline counterparts, attributed to an excess of low-frequency modes beyond the Debye prediction~\cite{anderson1972anomalous,argon1979plastic,buchenau1984neutron,falk1998dynamics,pohl2002low,widmer2008irreversible,xu2010anharmonic,chen2011measurement,manning2011vibrational,paoluzzi2019relation}. The intrinsic structural disorder in these solids gives rise to low-frequency modes localized in space, termed quasi-localized excitations (QLEs), which coexist with phonons. The displacements of particles in QLEs decay as $1/r^{d-1}$ in $d$ dimensions, where $r$ represents the distance from the core of the vibration~\cite{schober1991localized, lerner2021low}. Recent numerical studies consistently identify a $D(\omega) \sim \omega^4$ scaling behavior in various models of amorphous solids~\cite{lerner2016statistics,angelani2018probing, mizuno2017continuum, das2020robustness, bonfanti2020universal,richard2020universality,arceri2020vibrational,lerner2021low}. This quartic law has been observed in simulated glasses across different dimensions~\cite{ kapteijns2018universal}, as well as in glasses obtained through different annealing protocols~\cite{wang2019low}. Several mean-field arguments~\cite{galperin1989localized, gurarie2003bosonic} and phenomenological studies~\cite{gurevich2003anharmonicity, parshin2007vibrational} support the quartic scaling $D(\omega) \sim \omega^4$ for the distribution of low-frequency QLEs in disordered solids. However, effective medium theories, such as the ``Fluctuating Elasticity Theory'', propose deviations from the $D(\omega) \sim \omega^4$ law~\cite{schirmacher2011some,marruzzo2013heterogeneous,schirmacher2024nature}. While various aspects of the quartic law have been tested in simulations of amorphous solids, several studies have also observed deviations from this behavior. Poor annealing and small system sizes have been shown to yield power-law exponents less than 4~\cite{lerner2017effect, lerner2020finite, lerner2022nonphononic}. Recent studies have reported exponents of 3 and 3.5 in the low-frequency regime for two and three dimensions, respectively~\cite{wang2021low, wang2023scaling, wang2022density}. Additionally, confined three-dimensional thin films exhibit a low-frequency vibrational density of states following $D(\omega) \sim \omega^3$~\cite{yu2022omega}.

Numerical studies of amorphous solids often employ periodic simulation boxes to investigate bulk behaviors. While this approach is useful, it introduces residual shear stresses that can lead to solid arrangements that are unstable under boundary deformations~\cite{dagois2012soft,goodrich2014jamming}. Although the influence of boundary conditions has been studied for various properties of amorphous solids, their effect on the low-frequency vibrational spectrum remains relatively underexplored. Recently, it has been shown that periodic solids strained to their simple shear minima, where they are inherently stable to simple shear deformation, exhibit $D(\omega) \sim \omega^5$ in their low-frequency VDoS in both two and three-dimensional systems~\cite{krishnan2022universal}. Additionally, solids prepared under open boundary conditions, which allow relaxation in all degrees of freedom and stability under all perturbations, contain fewer low-frequency quasi-localized modes compared to those with periodic boundary conditions~\cite{chakraborty2023enhanced}. This is reflected in the vibrational spectrum, where $D(\omega) \sim \omega^4$ changes to $D(\omega)\sim \omega^\delta$ with $\delta=5$ in two dimensions and $\delta=4.5$ in three dimensions. Notably, modes localized near the surfaces of these solids tend to be softer, leading to a reduction in the scaling exponent for the VDoS. A recent study revealed distinct scaling behaviors in the distribution of low-frequency modes between solid configurations that are stable under boundary deformation and those that are unstable~\cite{xu2024low}. Specifically, configurations unstable to deformation exhibit a power-law scaling of $D(\omega) \sim \omega^{3.3}$, independent of dimensionality. In contrast, stable configurations show a dimension-dependent scaling, with $D(\omega) \sim \omega^{5.5}$ in two dimensions and $D(\omega) \sim \omega^{6.5}$ in three dimensions. The stability of these configurations against boundary deformations is indicated by a positive value of the lowest eigenvalue of the `Extended Hessian' (EH) matrix, which accounts for contributions of boundary degrees of freedom~\cite{goodrich2014jamming}.

\begin{figure}[t!]
\centering
\includegraphics[width=0.450\textwidth]{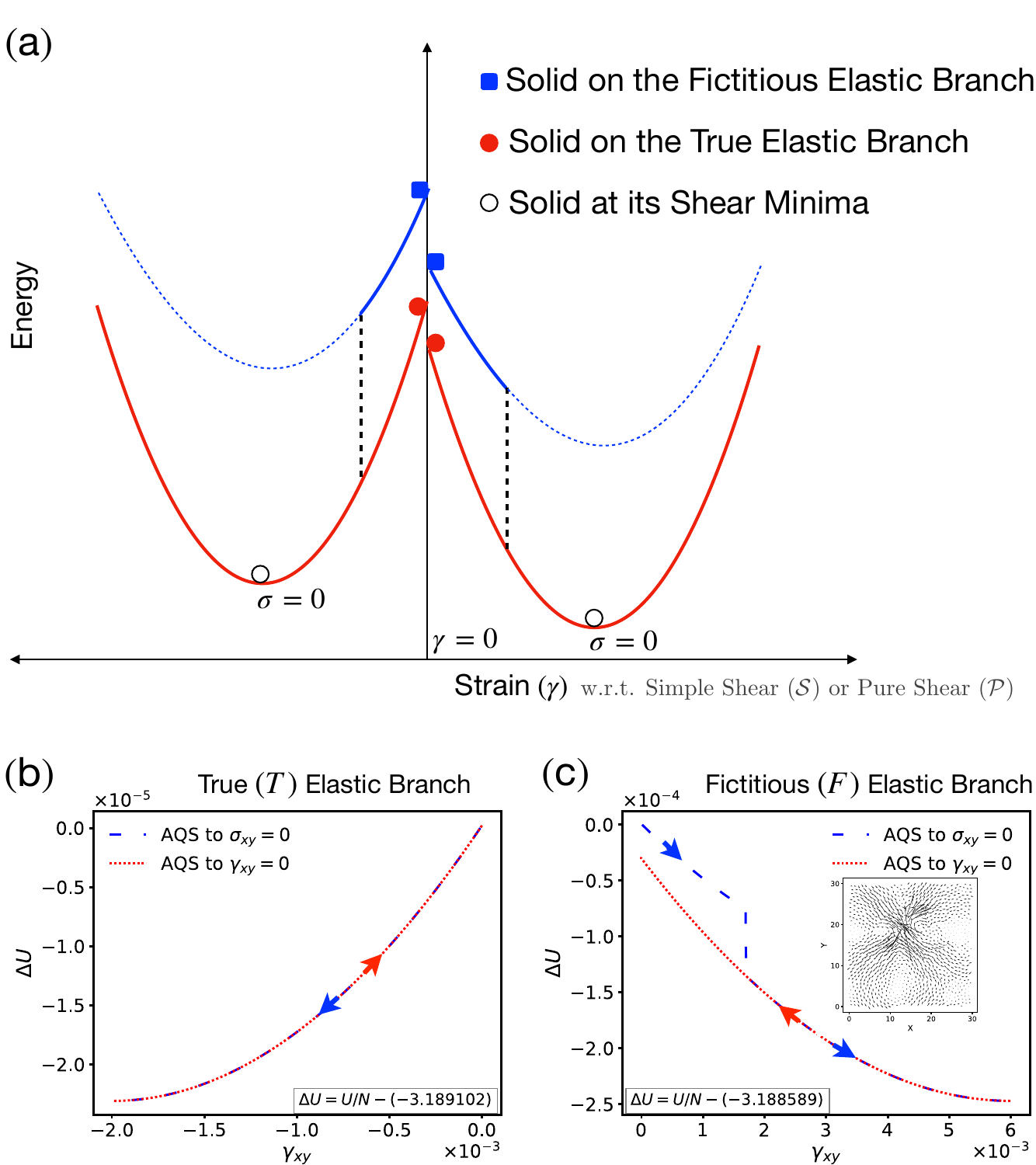}
\caption{(\textbf{a}) Schematic diagram illustrating the behavior of amorphous solids generated using standard simulation protocols with periodic boundary conditions. These boundary conditions often lead to solids residing on Fictitious elastic branches (\textbf{blue}), where shear minima are inaccessible through elastic deformation alone and undergo irreversible plastic events while straining towards the minima. Solids on the True elastic branch (\textbf{red}) can access shear minima via elastic deformations alone. Configurations at shear minima with zero residual shear stress are shown as open circles. Panels (\textbf{b}) and (\textbf{c}) display the energy as a function of strain in the cyclic AQS protocol for a typical configuration in two dimensions with a system size $N=1024$ residing on the True and Fictitious elastic branch under simple shear deformations, respectively. The inset in (\textbf{c}) illustrates the nature of the plastic event when transitioning from the Fictitious to the True elastic branch.
}
\label{landscape}
\end{figure}

\begin{table}[t!]
\centering
\resizebox{\columnwidth}{!}{
\begin{tabular}{|c|c|c|c|}
\hline
\textbf{Simple shear} & \textbf{Pure shear} & \textbf{Ensemble Classification} & \textbf{VDoS Power-Law} \\
\hline
\hline
\rule{0pt}{3ex}
\textbf{Fictitious} & \textbf{Fictitious} &  \textbf{\(\mathcal{S}F\)} + \textbf{\(\mathcal{P}F\)} & $\omega^{3}$ \\
\hline
\textbf{Fictitious} & \textbf{Either} &  \textbf{\(\mathcal{S}F\)} & $\omega^{3.5}$ \\
\hline
\textbf{Either} & \textbf{Fictitious} &  \textbf{\(\mathcal{P}F\)} & $\omega^{3.5}$ \\
\hline
\textbf{True} & \textbf{Fictitious} &  \textbf{\(\mathcal{S}T\)} + \textbf{\(\mathcal{P}F\)} & $\omega^{4.5}$ \\
\hline
\textbf{Fictitious} & \textbf{True} &  \textbf{\(\mathcal{S}F\)} + \textbf{\(\mathcal{P}T\)} & $\omega^{4.5}$ \\
\hline
\textbf{True} & \textbf{Either} &  \textbf{\(\mathcal{S}T\)} & $\omega^{5}$ \\
\hline
\textbf{Either} & \textbf{True} &  \textbf{\(\mathcal{P}T\)} & $\omega^{5}$ \\
\hline
\textbf{True} & \textbf{True} &  \textbf{\(\mathcal{S}T\)} + \textbf{\(\mathcal{P}T\)} & $\omega^{5.5}$\\
\hline
\textbf{Minima} & \textbf{Either} &  \textbf{\(\mathcal{S}_{min}\)} & $\omega^{5}$ \\
\hline
\textbf{Either} & \textbf{Minima} &  \textbf{\(\mathcal{P}_{min}\)} & $\omega^{5}$ \\
\hline
\textbf{Minima} & \textbf{Minima} &  \textbf{\(\mathcal{S}_{min}\)} + \textbf{\(\mathcal{P}_{min}\)} & $\omega^{6.5}$ \\
\hline
\end{tabular}
}
\caption{Power-law exponents for different ensembles of amorphous solids. Solids classified as \textbf{\(\mathcal{S}F\)} are on the Fictitious ($F$) elastic branch under simple (\textbf{\(\mathcal{S}\)}) shear deformations, where shear minima cannot be reached through elastic deformation alone and require a plastic event to transition to a lower-energy elastic branch. Similarly, solids on the Fictitious elastic branch under pure (\textbf{\(\mathcal{P}\)}) shear deformations are classified as \textbf{\(\mathcal{P}F\)}. Solids on the True (T) elastic branch, where shear minima can be reached via elastic deformation, are classified as \textbf{\(\mathcal{S}T\)} and \textbf{\(\mathcal{P}T\)} for simple and pure shear deformations, respectively. \textbf{\(\mathcal{S}\)$_{min}$} and \textbf{\(\mathcal{P}\)$_{min}$} denote solids at their simple and pure shear minima, respectively, exhibiting zero residual shear stresses.
}
\label{tab:power_laws}
\end{table}

In this study, we investigate the influence of shear instabilities on the vibrational spectrum of amorphous solids. The main contribution of this study is the identification of two distinct elastic branches - True ($T$) and Fictitious ($F$) that arise due to the effect of boundary conditions and differ in the nature of their energy landscapes. These branches significantly influence vibrational stability. Configurations generated under periodic boundary conditions often reside on Fictitious elastic branches, where shear minima cannot be accessed through elastic deformation alone. Instead, plastic instabilities are necessary to transition the system to a lower-energy elastic branch (see Fig.~\ref{landscape} (\textbf{a})). By applying zero-temperature simple and pure shear deformations, we identify these states and observe that their low-frequency VDoS follows a power-law distribution, $D(\omega) \sim \omega^\delta$, with $\delta \approx 3$. In contrast, configurations on True elastic branches, where shear minima are accessible through elastic deformation, exhibit a lower propensity for quasi-localized excitations. These vibrationally stable solids display a low-frequency VDoS scaling as $D(\omega) \sim \omega^{5.5}$. When considering all configurations together, the ensemble-averaged distribution conforms to $D(\omega) \sim \omega^4$. Importantly, solids on True elastic branches may undergo plastic instabilities when deformed away from their shear minima, potentially appearing unstable in an EH analysis. Conversely, solids that appear stable with an EH analysis can still reside on Fictitious branches, as EH identifies configurations prone to plastic instability only within a limited range of deformations. A steeper power-law behavior, $D(\omega) \sim \omega^{6.5}$, is observed for solids that are both EH-stable and reside on True elastic branches. Additionally, solids fully relaxed to their shear minima, with no residual shear stress, stable under all deformations, and residing on True elastic branches are characterized by the same scaling behaviour. Interestingly, while the fraction of solids unstable to boundary deformations within the EH framework decreases with system size, the prevalence of Fictitious branch states increases. This leads to two limiting scenarios for the low-frequency VDoS: If system size is increased without addressing instabilities, the VDoS will scale as $D(\omega) \sim \omega^3$. Conversely, by removing residual shear stress before increasing system size, we expect the VDoS to scale as $D(\omega) \sim \omega^{6.5}$ across both two and three dimensions.
\begin{figure*}
\centering
\includegraphics[width=0.950\textwidth]{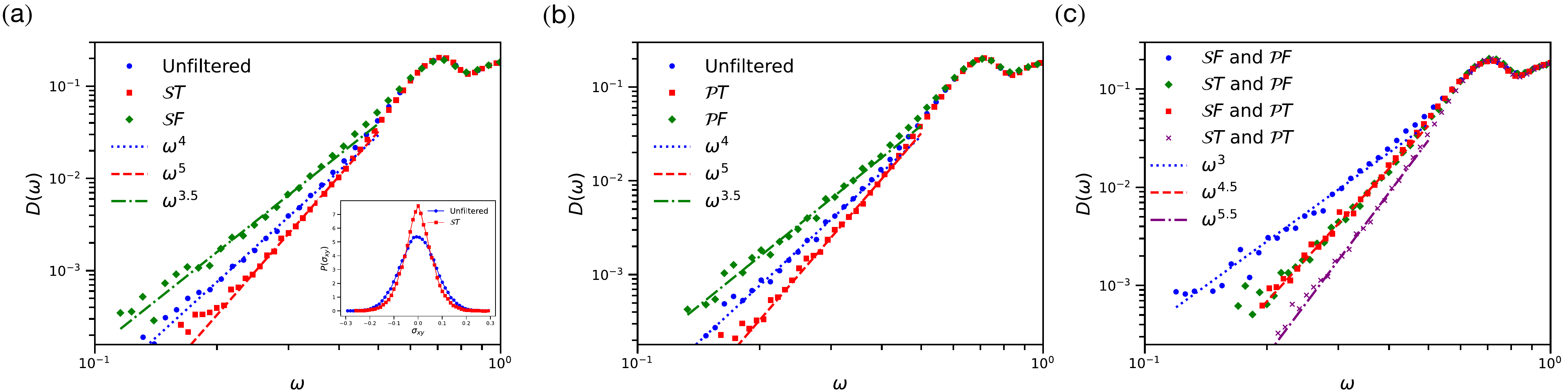}
\caption{Low-frequency vibrational density of states (VDoS) for two-dimensional, amorphous solids with system size \(N=1024\), categorized by elastic branch type under simple and pure shear deformations. (\textbf{a}) VDoS for solids under simple shear deformation. Configurations classified as \textbf{\(\mathcal{S}F\)} are on Fictitious elastic branches and exhibit $D(\omega) \sim \omega^{3.5}$, indicating a high density of low-frequency modes. In contrast, configurations on the True elastic branch (\textbf{\(\mathcal{S}T\)}) show a more stable $D(\omega) \sim \omega^5$ behavior at their initial state ($\gamma_{xy}=0$), with non-zero residual shear stresses (inset). When combining \textbf{\(\mathcal{S}T\)} and \textbf{\(\mathcal{S}F\)} solids, the resulting VDoS exhibits a scaling behavior of $D(\omega) \sim \omega^4$. (\textbf{b}) VDoS for solids under pure shear deformation. Similar to simple shear results, solids on Fictitious elastic branches (\textbf{\(\mathcal{P}F\)}) exhibit $D(\omega) \sim \omega^{3.5}$, while those on True branches (\textbf{\(\mathcal{P}T\)}) show $D(\omega) \sim \omega^5$. The ensemble of \textbf{\(\mathcal{P}T\)} and \textbf{\(\mathcal{P}F\)} configurations display $D(\omega) \sim \omega^4$ behavior. (\textbf{c}) VDoS for ensembles combining solids from different elastic branches under both shear deformations. Solids that reside on Fictitious elastic branches in both types of shear deformations (\textbf{\(\mathcal{S}F\)} and \textbf{\(\mathcal{P}F\)}) exhibit a pronounced low-frequency mode density with a scaling behavior of $D(\omega) \sim \omega^3$. Intermediate scaling of $D(\omega) \sim \omega^{4.5}$ is observed in ensembles of \textbf{\(\mathcal{S}T\)} and \textbf{\(\mathcal{P}F\)}, or \textbf{\(\mathcal{S}F\)} and \textbf{\(\mathcal{P}T\)}. Solids that are on True elastic branches under both shear deformations (\textbf{\(\mathcal{S}T\)} and \textbf{\(\mathcal{P}T\)}) exhibit a steeper power-law $D(\omega) \sim \omega^{5.5}$, indicating a significant suppression of low-frequency modes and increased stability.}
\label{2D_VDoS}
\end{figure*}

{\large\textbf{Results}}

We study the canonical Kob-Andersen Lennard-Jones mixture~\cite{kob1995testing} in both two-dimensional (2D) and three-dimensional (3D) systems. The model consists of binary particle mixtures with a composition ratio of $65:35$ and a number density of $1.15$ in 2D, and a composition ratio of $80:20$ at zero pressure in 3D. The interaction potential between particle types is described by a modified Lennard-Jones potential, enhanced with additional polynomial terms to smooth the potential at the truncation distance (see Appendix~\ref{model} for details). Initially, we equilibrate the liquid states at a high temperature ($T = 0.55$ for both 2D and 3D) using constant temperature molecular dynamics simulations under periodic boundary conditions. Solid configurations are then generated via conjugate gradient minimization~\cite{shewchuk1994introduction}. Simulations of solids at their shear minima are performed using LAMMPS~\cite{thompson2022lammps}. Statistics for all system sizes are performed, generating at least $10^5$ configurations in the initial ensemble for both 2D and 3D.
Configurations are classified as follows: `\textbf{\(\mathcal{S}T\)}' for True elastic branches under simple shear, `\textbf{\(\mathcal{S}F\)}' for Fictitious elastic branches under simple shear, `\textbf{\(\mathcal{P}T\)}' and `\textbf{\(\mathcal{P}F\)}' for True and Fictitious elastic branches under pure shear, respectively, and `\textbf{\(\mathcal{S}\)$_{min}$}' and `\textbf{\(\mathcal{P}\)$_{min}$}' for solids strained to their simple and pure shear minima, respectively. The distribution of solids across these branches leads to variations in the power-law behavior of the ensemble-averaged low-frequency VDoS, as summarized in Table~\ref{tab:power_laws}.

\vspace{0.1cm}
\textbf{Classification of Elastic Branches}
\vspace{0.1cm}

\textit{Elastic Branches Under Simple Shear}:
Amorphous solids created under periodic boundary conditions contain residual shear stresses, leading to such solids being unstable to deformations~\cite{dagois2012soft,goodrich2012finite,xu2024low}. Recent studies suggest that changes in this macroscopic shear stress {\it modify} the low-frequency VDoS, with the power-law being modified to $D(\omega)\sim \omega^5$ when simple shear is applied to create a zero stress state~\cite{krishnan2022universal}. Although the residual shear stresses decrease with system size, the power-law observed in unstrained systems deviates from such a $D(\omega)\sim\omega^5$ behavior. These observations suggest a deeper underlying connection between the shear stability of amorphous solids and their low-frequency VDoS, which we explore in detail below.

We begin with amorphous solids in orthogonal periodic cells exhibiting residual shear stresses and apply simple shear deformations to reach the shear minima (where $\sigma_{xy} = 0$ in 2D). In each configuration, we measure the global shear stress ($\sigma_{xy}$) and incrementally strain the system towards the shear minima. This process involves applying small strain steps, followed by energy minimization using Lees-Edwards boundary conditions while adjusting the strain steps until the global shear stress magnitude falls below $10^{-8}$.

To classify configurations as either \textbf{\(\mathcal{S}T\)} or \textbf{\(\mathcal{S}F\)} under simple shear, we perform cyclic shear deformations that return the system to zero strain ($\gamma_{xy} = 0$) from the shear minimum. Configurations that undergo plastic events during the Athermal Quasi-Static (AQS) process to their shear minima, are identified by computing the mean squared displacement (MSD) between the initial and cyclically deformed configurations at $\gamma_{xy} = 0$, $\text{MSD} = \frac{1}{N} \sum_i^N |\vec{r}_f^{~i} - \vec{r}_0^{~i}|^2 $.
Here, $\vec{r}_0^{~i}$ and $\vec{r}_f^{~i}$ represent the position of the $i$-th particle in the initial and cyclically deformed configurations, respectively. Configurations on the True elastic branch, undergoing only reversible elastic events, exhibit $\text{MSD} = 0$ and are classified as \textbf{\(\mathcal{S}T\)}. In contrast, configurations that undergo irreversible plastic events, transitioning from Fictitious elastic branches to the True elastic branch, result in a finite MSD and are classified as \textbf{\(\mathcal{S}F\)} in their initial state at $\gamma = 0$. 
Fig.~\ref{landscape} (\textbf{b}) and (\textbf{c}) illustrate the energy per particle as a function of strain in the cyclic AQS protocol for typical \textbf{\(\mathcal{S}T\)} and \textbf{\(\mathcal{S}F\)} configurations, respectively, in a 2D system with $N=1024$. The inset in Fig.~\ref{landscape} (\textbf{c}) shows the plastic event associated with the transition from a Fictitious to a True elastic branch, with the displacement field displaying a characteristic quadrupolar pattern~\cite{lemaitre2009rate}.

We then filter out the \textbf{\(\mathcal{S}T\)} and \textbf{\(\mathcal{S}F\)} configurations from the initial ensemble. Fig.~\ref{2D_VDoS} (\textbf{a}) shows the low-frequency VDoS of these two classes of solids. Solids residing on the higher-energy Fictitious elastic branches \textbf{\(\mathcal{S}F\)} exhibit a lower scaling exponent, $D(\omega) \sim \omega^{3.5}$, indicative of a greater propensity for low-frequency vibrational modes. In contrast, configurations on the True elastic branch are more stable, following a $D(\omega) \sim \omega^5$ scaling, although they retain non-zero residual shear stresses (as shown in the inset of Fig.~\ref{2D_VDoS} (\textbf{a})). An ensemble combining both \textbf{\(\mathcal{S}T\)} and \textbf{\(\mathcal{S}F\)} solids results in $D(\omega) \sim \omega^4$. 

\textit{Elastic Branches Under Pure Shear}: We next investigate solids subjected to pure shear deformations, focusing on fluctuations in the diagonal stress components. Starting with constant pressure ($P$) configurations, which exhibit diagonal stress fluctuations such that $\sigma_{xx} + \sigma_{yy} = P$, resulting in a finite tensile stress component ($\sigma_{xx} - \sigma_{yy} \ne 0$). To control these fluctuations, we expand along the x-direction and compress along the y-direction (or vice versa) in small steps, followed by energy minimization to make each diagonal stress constant ($\sigma_{xx} = \sigma_{yy} = P$). This process maintains constant pressure while introducing slight changes in area (or volume). Although this differs from conventional pure shear, it can be considered a similar protocol.

Fig.~\ref{2D_VDoS} (\textbf{b}) illustrates the low-frequency VDoS based on response to pure shear deformations. Solids that do not incur plastic rearrangements while straining to the minima are classified as \textbf{\(\mathcal{P}T\)}, while solids that undergo plastic rearrangement (as indicated by a finite $\text{MSD}$ when we rescale the box dimensions to their initial values) are identified as \textbf{\(\mathcal{P}F\)}. We observe similar behavior in the low-frequency VDoS, solids in the \textbf{\(\mathcal{P}T\)} category show $D(\omega) \sim \omega^5$ behavior, whereas \textbf{\(\mathcal{P}F\)} solids show a lower exponent, $D(\omega) \sim \omega^{3.5}$ behavior. When both  \textbf{\(\mathcal{P}T\)} and  \textbf{\(\mathcal{P}F\)} configurations are considered together, the ensemble displays $D(\omega) \sim \omega^4$ behavior. 

\textit{Scaling Behavior Combining True and Fictitious Elastic Branches}: In the previously discussed ensembles, solids were categorized based on their response to either simple or pure shear deformations. Hence the classifications based on relaxation to shear minima under simple shear do not distinguish between configurations that may undergo plastic rearrangements under pure shear, and vice versa. Consequently, the vibrational spectra of the ((\textbf{\(\mathcal{S}T\)} and \textbf{\(\mathcal{S}F\)})) ensembles include contributions from configurations that may or may not undergo plastic rearrangements under pure shear. Similarly, the spectra of \textbf{\(\mathcal{P}T\)} and \textbf{\(\mathcal{P}F\)} reflect contributions from solids on different elastic branches under simple shear. We now examine ensembles categorized by stability under both shear deformations, defined through combinations of \textbf{\(\mathcal{S}T\)},  \textbf{\(\mathcal{S}F\)},  \textbf{\(\mathcal{P}T\)}, and  \textbf{\(\mathcal{P}F\)}. As illustrated in Fig.~\ref{2D_VDoS}(c), these ensembles display distinct scaling behaviors in the VDoS, revealing the influence of True and Fictitious elastic branches on the low-frequency vibrational spectrum under both simple and pure shear deformations.

Solids residing on Fictitious elastic branches in both shear deformations-\textbf{\(\mathcal{S}F\)} and  \textbf{\(\mathcal{P}F\)}-exhibit a large propensity of low-frequency modes, characterized by the lowest scaling exponent,
$D(\omega)\sim\omega^{3}$. This behavior suggests that the observed scaling of $D(\omega)\sim\omega^{3}$ in experiments on quasi-two dimensional confined systems may be attributed to the prevalence of solids on Fictitious elastic branches resulting due to confinement effects~\cite{yu2022omega}. Moreover, the exponent close to $3$ reported in large ensembles of two- and three-dimensional amorphous systems may arise from a significant fraction of configurations residing on Fictitious branches at the system sizes studied~\cite{wang2021low, wang2022density, wang2023scaling}.

In contrast, ensembles containing solids that reside on the True elastic branch under either simple shear or pure shear, while being on a Fictitious branch under the other deformation, exhibit intermediate VDoS behavior. These configurations-\textbf{\(\mathcal{S}T\)} and \textbf{\(\mathcal{P}F\)}, or  \textbf{\(\mathcal{S}F\)} and  \textbf{\(\mathcal{P}T\)}-display a scaling of  $\omega^{4.5}$. Finally, we consider the ensemble of solids classified as \textbf{\(\mathcal{S}T\)} and  \textbf{\(\mathcal{P}T\)}. This ensemble represents configurations that reside on the True elastic branches under both simple shear and pure shear deformations. These solids exhibit a scaling behavior of $D(\omega)\sim\omega^{5.5}$, indicating a significant suppression of low-frequency modes, reflecting the enhanced stability of these solids. We examined this power-law behavior by rescaling the box lengths and applying additional strain in the simple shear direction. Solids on the True elastic branches in both simple and pure shear are characterized by shear minima that are accessible purely through elastic deformation across all shear directions and show zero mean-squared displacement (MSD = 0) between the initial configurations and those obtained through reverse deformations, as discussed in Appendix~\ref{both_deformation}.

Importantly, the ensembles we have discussed so far are characterized by whether the solids reside on True elastic branches or Fictitious elastic branches. This classification is determined by the presence of shear minima or the occurrence of plastic jumps to lower energy branches. Solids on the True elastic branch may still have nearby plastic instabilities when deformed in the opposite directions. Thus, these ensembles include solids that are prone to plastic instabilities when deformed away from their shear minima.
\begin{figure}
\centering
\includegraphics[width=0.450\textwidth]{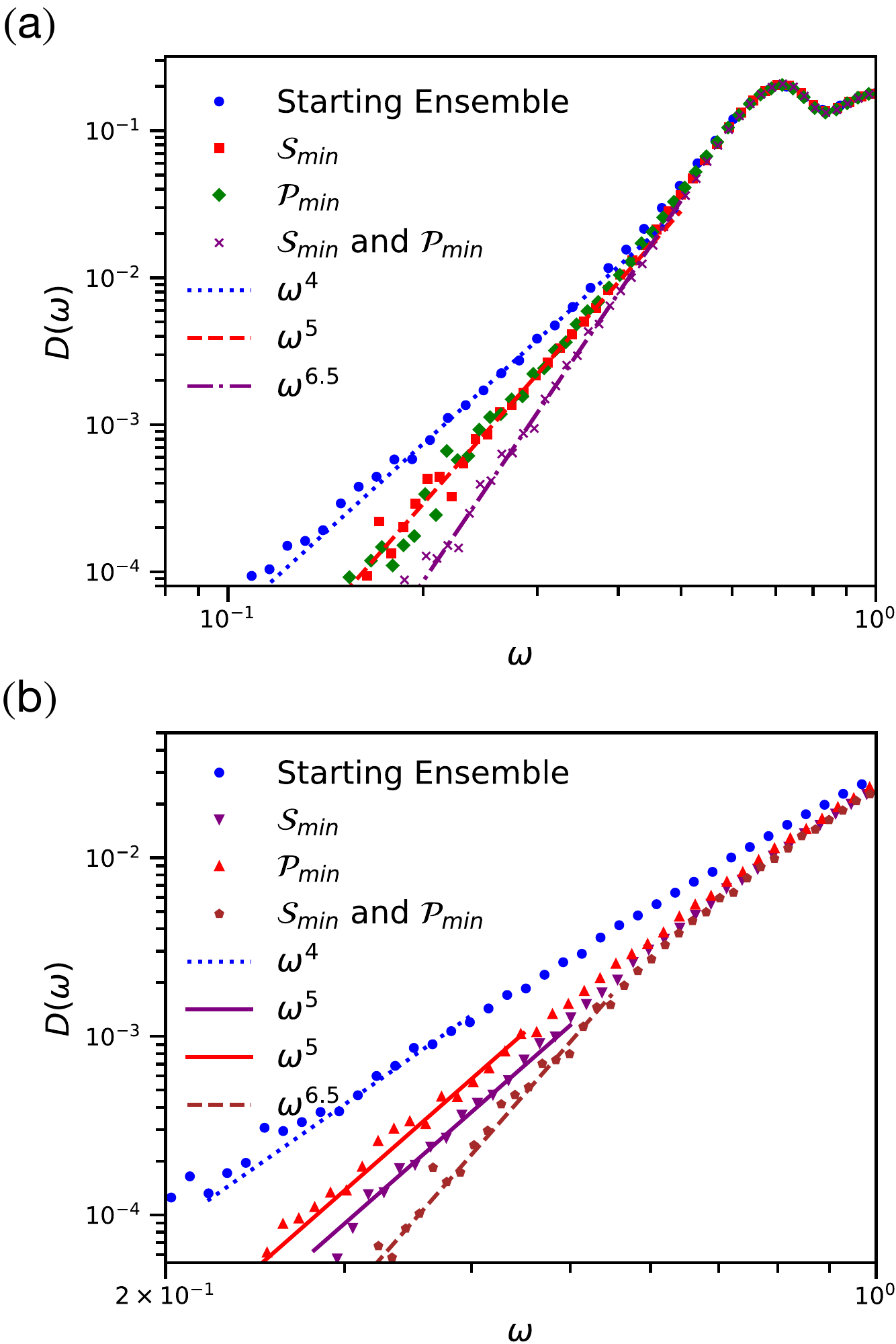}
\caption{VDoS for amorphous solids with zero residual shear stresses. \textbf{(a)} Two-dimensional solids with $N = 1024$ particles. For solids at the minima of either simple or pure shear deformations, denoted as \textbf{\(\mathcal{S}\)$_{min}$} or \textbf{\(\mathcal{P}\)$_{min}$}, the VDoS follows a scaling of $D(\omega) \sim \omega^{5}$, indicative of solids on a True elastic branch in any single shear direction. Solids achieving shear minima across all directions, where all residual shear stress components are zero, exhibit even greater stability with a VDoS scaling of $D(\omega) \sim \omega^{6.5}$. This suggests that complete relaxation under various shear deformations results in a more stable state, characterized by a higher exponent in the VDoS scaling. \textbf{(b)} Three-dimensional solids with $N = 1000$ particles. In configurations where all macroscopic stress components fluctuate, leading to Fictitious elastic branches, the VDoS scales as $D(\omega) \sim \omega^{4}$. When all simple shear components are constrained to zero, the VDoS scales as $D(\omega) \sim \omega^{5}$. The same scaling is observed when all pure shear components are constrained to zero. The VDoS follows a scaling of $D(\omega) \sim \omega^{6.5}$ for ensembles of solids stable under all shear deformations.}
\label{lammps_vdos}
\end{figure}

\vspace{0.1cm}
\textbf{Stable Configurations with Zero Residual Shear Stresses}
\vspace{0.1cm}

Next, we examine the low-frequency vibrational spectrum of solids residing at their shear minima, denoted as \textbf{\(\mathcal{S}\)$_{min}$} for simple shear and \textbf{\(\mathcal{P}\)$_{min}$} for pure shear. At these minima, the solids exhibit no residual tensile stress components, indicating that the system has fully relaxed under the applied deformation. Consequently, they are stable under all perturbations and reside on the true elastic branches of all shear degrees of freedom.

To achieve configurations at shear minima, which are also energy minimized in the position degrees of freedom, we deform the periodic cell during energy minimization. In $d$ dimensions, there are $ \frac{d(d+1)}{2} - 1 $ independent shear directions. In two dimensions, this includes one simple shear direction and one pure shear direction. In three dimensions, there are five such shear directions: three simple shear directions ($\sigma_{xy}$, $\sigma_{yz}$, and $\sigma_{zx}$) and two pure shear components ($\sigma_{xx} - \sigma_{yy}$) and ($\sigma_{xx} + \sigma_{yy} - 2\sigma_{zz}$). To achieve configurations where all shear stress components are minimized, we rescale the box lengths and apply additional straining in the simple shear direction, iterating this process until fluctuations in all stress components fall below $10^{-8}$. Simulations are performed using LAMMPS~\cite{thompson2022lammps}, with macroscopic stress components controlled via the box/relax' command~\cite{lammps}, as described in Appendix~\ref{model}.

Fig.~\ref{lammps_vdos} \textbf{(a)} shows the low-frequency VDoS for two-dimensional solids with $N = 1024$ particles. Solids at the minima of either simple or pure shear deformations \textbf{\(\mathcal{S}\)$_{min}$} or \textbf{\(\mathcal{P}\)$_{min}$} exhibit a VDoS scaling of $D(\omega) \sim \omega^{5}$, characteristic of solids on a True elastic branch in any single shear direction. Interestingly, when solids reach shear minima across all shear directions, i.e., all residual shear stress components are zero, they exhibit even greater stability, with the VDoS scaling as $D(\omega) \sim \omega^{6.5}$. This suggests that complete relaxation under various {\it shear} deformations leads to a more {\it vibrationally} stable state, characterized by a higher exponent in the VDoS scaling.
Fig.~\ref{lammps_vdos} \textbf{(b)} presents the low-frequency VDoS for three-dimensional solids with $N = 1000$ particles. When all macroscopic stress components are allowed to fluctuate, resulting in configurations on Fictitious elastic branches, the VDoS follows a $ D(\omega) \sim \omega^4 $ power law. When all simple shear components are constrained to zero, the VDoS scales as $ D(\omega) \sim \omega^5 $. Similar scaling is observed when all pure shear components are constrained to zero. In ensembles of solids at minima for all shear directions, and thus stable under all shear deformations, the VDoS exhibits a scaling of $ D(\omega) \sim \omega^{6.5}$.

At this point, we would like to compare and contrast our findings with a recent study by Xu {\it et al.}~\cite{xu2024low}, where stable two-dimensional solids were shown to exhibit an $\omega^{5.5}$ scaling behavior, while stable three-dimensional solids demonstrated an $\omega^{6.5}$ scaling behavior. In their study, the distinction between stable and unstable configurations was made by examining the eigenvalues of the EH, which accounts for boundary deformations (discussed in Appendix~\ref{appen_EH}). Such an EH analysis explores small domains in boundary deformations, identifying configurations close to plastic instabilities as unstable. As discussed, the ensembles we previously considered may have plastic instabilities when deformed away from their shear minima. Thus, solids on True elastic branches may appear unstable within such an EH analysis, as shown in Appendix~\ref{appen_EH}. Similarly, the EH analysis identifies stable solids from both True and Fictitious elastic branches, as the analysis is conducted within a restricted domain. For the two-dimensional solids in our study, our EH analysis identified a fraction $9 \times 10^{-3}$ of configurations as unstable, and after removing them from the ensemble, these EH-stable solids exhibit a $D(\omega) \sim \omega^{5.5}$ behaviour. 
\begin{figure}
\centering
\includegraphics[width=0.49\textwidth]{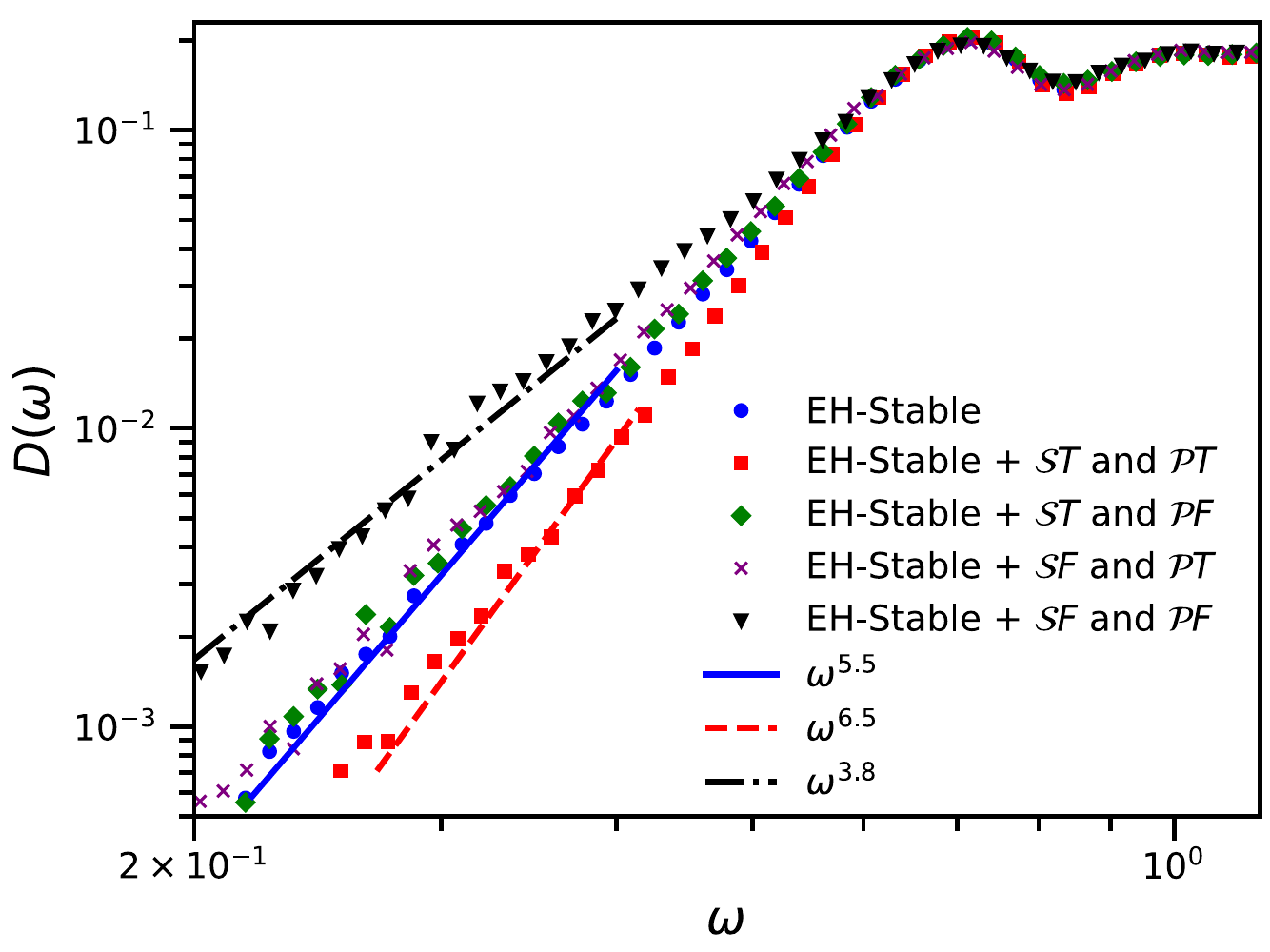}
\caption{Low frequency VDoS for solids stable under EH analysis, categorized by their respective elastic branches. EH-stable solids residing on the True elastic branch for both simple and pure shear deformations exhibit a steeper power-law behavior with $D(\omega) \sim \omega^{6.5}$, similar to the scaling behavior observed in solids at shear minima. In contrast, those on Fictitious branches demonstrate power-law scaling with exponents less than $5.5$. This suggests that while these configurations are stable according to EH analysis, they do not posses the same degree of vibrational stability as those on the True branches. Considering solids from both branches, the overall scaling behavior for the EH-stable ensemble is characterized by $D(\omega) \sim \omega^{5.5}$.}
\label{VDoS_TF_EH}
\end{figure}
Among these EH-stable configurations, roughly half of them reside on Fictitious branches of either simple or pure shear. Grouping EH-stable configurations by their respective elastic branches, we observe that solids stable under an EH analysis in True elastic branches exhibit an $\omega^{6.5}$ scaling in their low-frequency VDoS, as illustrated in Fig.~\ref{VDoS_TF_EH}. In contrast, EH-stable solids on Fictitious elastic branches result in a scaling exponent less than $5.5$. When considering all EH-stable solids together, the overall scaling behavior tends to $\omega^{5.5}$. This leads us to speculate that in three dimensions, where more shear directions are present, EH-stable solids that are not prone to plastic instabilities are likely to be on True elastic branches for both simple and pure shear directions. This is supported by the observed $\omega^{6.5}$ scaling of solids residing at the shear minima in both two and three dimensions ($\mathcal{S}_{min}$ and $\mathcal{P}_{min}$). Solids with residual shear stress equal to zero, which are stable under all deformations and reside on True elastic branches, exhibit an $\omega^{6.5}$ behavior across both two and three dimensions.

\vspace{0.1cm}
\textbf{System Size Dependence}
\vspace{0.1cm}

The distribution of low-frequency vibrations in amorphous solids exhibits finite-size effects. In very small systems, an excess of quasilocalized vibrations is characterized by scaling behavior $D(\omega) \sim \omega^{\delta}$, where $\delta < 4$~\cite{lerner2020finite, xu2024low}, likely due to a higher prevalence of EH-unstable configurations. As system size increases, the fraction of EH-unstable configurations decreases, which is expected to result in a vibrational spectrum more characteristic of stable solids. However, even in EH-stable solids, the regime characterized by $D(\omega)\sim\omega^{6.5}$ becomes suppressed in three-dimensional systems as size increases, gradually transitioning towards $\omega^{4}$ behavior~\cite{xu2024low}. Furthermore, studies have reported exponents below $4$ in large two- and three-dimensional ensembles at system sizes where EH-instabilities are expected to be negligible~\cite{wang2021low, wang2022density, wang2023scaling}.
\begin{figure}
\centering
\includegraphics[width=0.45\textwidth]{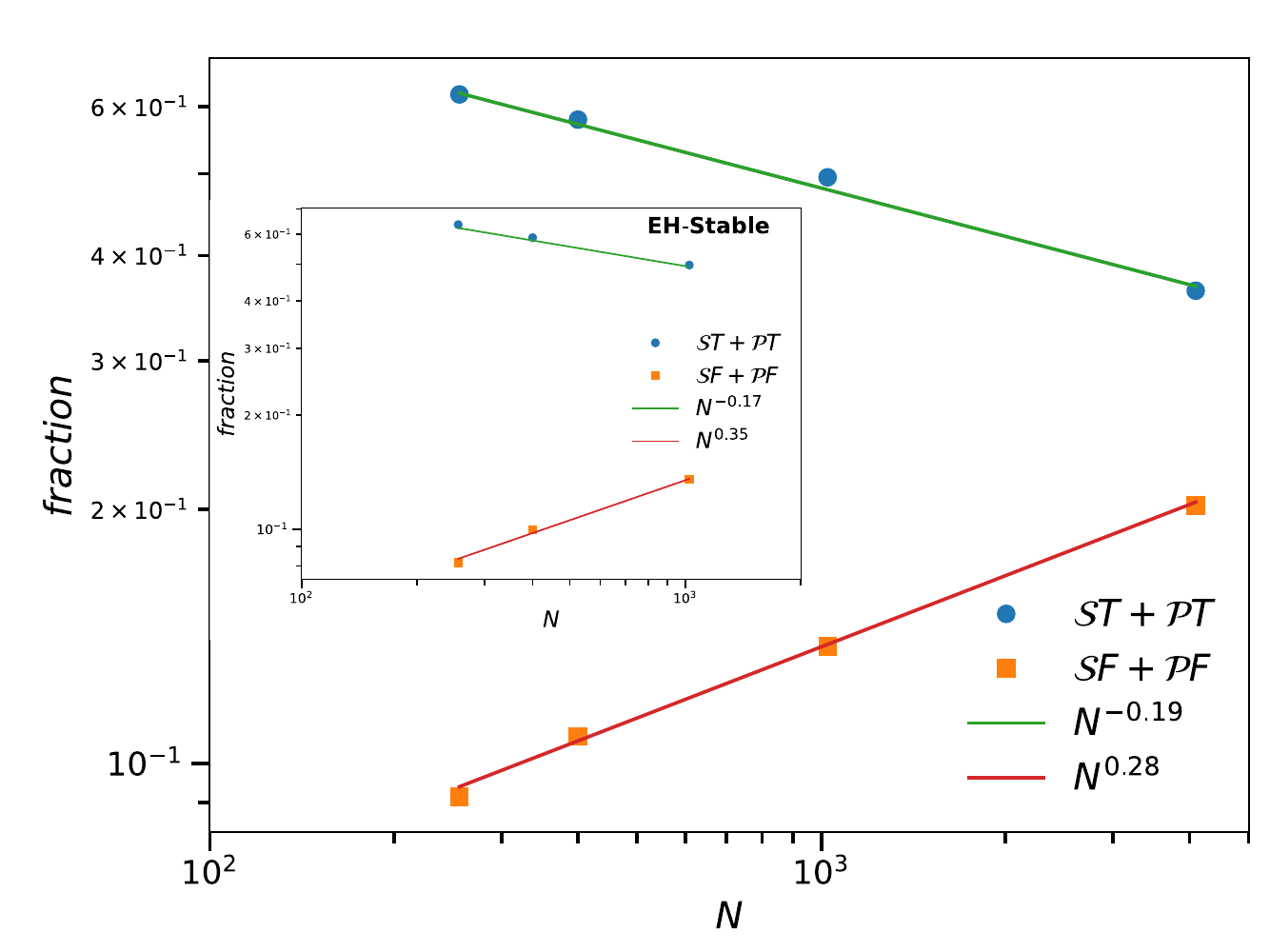}
\caption{System size dependence of two-dimensional solids residing on True and Fictitious elastic branches under simple and pure shear deformations. As system size increases, the fraction of solids residing on Fictitious branches, where shear minima are inaccessible via elastic deformations, also increases. This suggests that with increasing system size, even infinitesimal residual shear stress leads to a larger proportion of solids on Fictitious branches. To generate solids on True elastic branches, the limit of zero residual shear stress must be taken first. Inset shows the behavior of solids stable under an EH analysis.}
\label{fraction}
\end{figure}

To explore these effects, we investigate how system size influences the fraction of solids residing on True and Fictitious elastic branches in two-dimensional solids. Fig.~\ref{fraction} illustrates the dependence of these solids on elastic branches under both simple and pure shear deformations. As system size increases, the fraction of solids on Fictitious elastic branches-where shear minima cannot be achieved through elastic deformations and require plastic instabilities for transitions to lower energy states-also increases, despite a decrease in residual shear stresses. The inset of Fig.~\ref{fraction} depicts this behavior for solids that are stable under the EH analysis. Notably, with increasing system size, the fraction of solids classified as unstable under the EH analysis and prone to plastic instabilities decreases~\cite{xu2024low}. This observation suggests a significant shift in the nature of instabilities as system size increases. Specifically, instabilities associated with Fictitious elastic branches begin to dominate the low-frequency vibrational spectrum in the ensemble-averaged distribution. Beyond a characteristic system size (approximately $N \sim 10^4$ particles for this model in two dimensions), the power-law exponent is expected to approach a value near 3 due to the substantial fraction of solids on the Fictitious elastic branch. However, at larger system sizes, phononic modes become increasingly dominant, complicating the detection of QLEs. As a result, a significant number of configurations are required to achieve good statistics and a clear observation of the power-law behavior.

\begin{figure}
\centering
\includegraphics[width=0.45\textwidth]{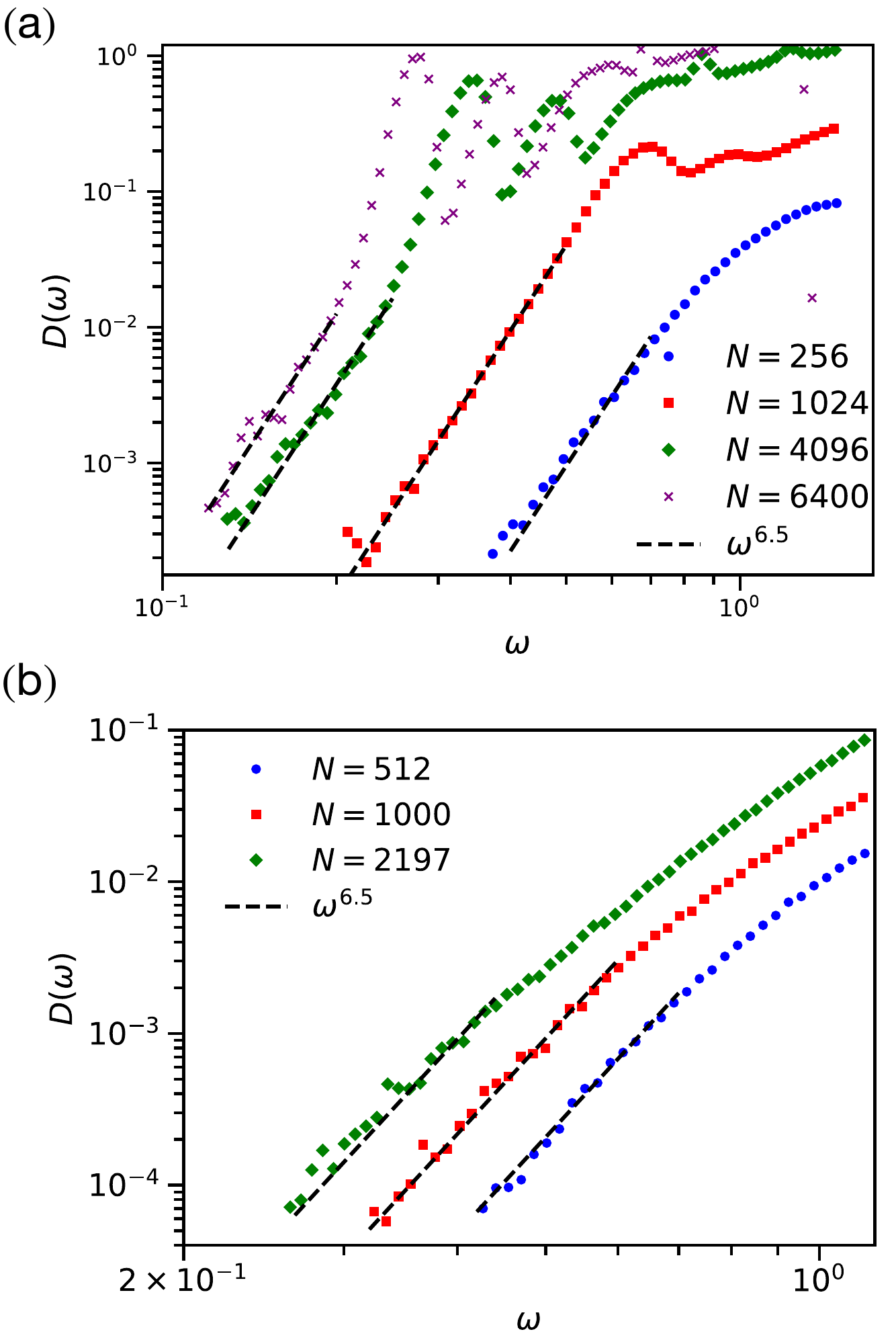}
\caption{
System size analysis of low-frequency VDoS scaling behavior for solids stable under all shear deformations. \textbf{(a)} Two-dimensional systems and \textbf{(b)} three-dimensional systems, both exhibiting a VDoS scaling of $D(\omega) \sim \omega^{6.5}$ across various system sizes. 
}
\label{system_size}
\end{figure}
Next, we conduct a system size analysis of the low-frequency VDoS for solids at their shear minima in both two and three dimensions. In ensembles of solids that are at the minimum of {\it all} shear directions and thus stable under all shear deformations, the VDoS exhibits a scaling behavior of $ D(\omega) \sim \omega^{6.5} $. Fig.~\ref{system_size} \textbf{(a)} and \textbf{(b)} illustrates the $ D(\omega) \sim \omega^{6.5} $ scaling of the low-frequency VDoS for various system sizes in both two and three dimensions.

These observations have significant implications for real stable solids. If we consider the limit $N \to \infty$ with all instabilities present, the observed power law will approach an exponent close to $3$. This is because, as the system size increases, the propensity for solids to reside on Fictitious elastic branches increases despite decreasing residual shear stress. Even small residual shear stress can lead to Fictitious elastic branches in large systems. Conversely, if we first take the limit of zero residual shear stress and then $N \to \infty$, we expect to observe $\omega^{6.5}$ behavior in the low-frequency VDoS of amorphous solids in both two and three dimensions.

{\large\textbf{Discussion}}

In summary, our study contributes to the emerging understanding of the distribution of non-phononic vibrations in amorphous solids. While previous research has suggested a $D(\omega) \sim \omega^4$ distribution for low-frequency vibrations, independent of interatomic interactions, thermal history, and spatial dimensionality, the exact scaling exponent remains debated. 
In this study, we have explored the critical role of boundary-condition-induced instabilities on the vibrational spectrum of amorphous solids. Our findings reveal the existence of two distinct elastic branches-True and Fictitious- that significantly impact the scaling behavior of the vibrational density of states at low frequencies. Configurations on Fictitious branches, where shear minima cannot be reached through elastic deformations alone and plastic instabilities are required to relax these configurations to their shear minima, exhibit a scaling of $D(\omega) \sim \omega^3$. The prevalence of such configurations increases as the system size grows. In contrast, solids residing on True elastic branches, where shear minima are accessible via elastic deformations, display a more stable vibrational spectrum, scaling as $D(\omega) \sim \omega^{5.5}$. Ensemble averaging across both types of branches leads to the frequently observed $D(\omega) \sim \omega^4$ behavior. Moreover, we show that solids relaxed to their shear minima, with no residual shear stress, exhibit the most stable vibrational spectra, characterized by $D(\omega) \sim \omega^{6.5}$ in both two and three dimensions. This same scaling is also observed in solids on True elastic branches that are not prone to plastic instabilities under small external perturbations. These findings emphasize the importance of controlling boundary conditions and residual stresses in simulations and experiments to study amorphous materials' vibrational properties .

In conclusion, we propose that the scaling behavior of low-frequency non-phononic vibrational modes in amorphous solids, arises from the aggregation of solids belonging to different types of elastic branches, with differing limiting behaviors in the $N \to \infty$ limit. If instabilities are not addressed, the VDoS approaches a $\omega^3$ scaling in large systems. However, by first eliminating residual shear stress before taking the large system size limit, we predict a $\omega^{6.5}$ scaling for the low-frequency in both two- and three-dimensional amorphous solids. Our findings  also have measurable experimental consequences, as controlled shear deformations of amorphous materials can be used to achieve a relaxation to their respective shear minima. 
The changes produced in the  vibrational density of states can, therefore, be extracted using inelastic neutron scattering, as has been routinely used to study the boson peak in glasses~\cite{chumakov2004collective,yu2022omega}. Recent advances in strain control and energy minimization techniques~\cite{zhang2010statistical,ren2013reynolds} therefore make it increasingly feasible to verify the different scaling behaviors in the VDoS of stable amorphous solids.


\textbf{Acknowledgments:} We thank Vishnu V. Krishnan, Edan Lerner and Eran Bouchbinder for useful discussions.  This project was funded by intramural funds at TIFR Hyderabad under Project Identification No. RTI 4007 from the Department of Atomic Energy, Government of India. S. K. would like to acknowledge support through the Swarna Jayanti Fellowship Grant No. DST/SJF/PSA-01/2018-19 and SB/SFJ/2019-20/05. The work of K.~R. was partially supported by the SERB-MATRICS grant MTR/2022/000966.

\textbf{Author contributions:} K.R. and S.K. designed the study. S.C. performed the numerical simulations and data analysis. R.M. conducted additional data analysis. All authors contributed to the writing of the manuscript.

\begin{appendix}

\section{Model}\label{model}
We study the canonical Kob-Andersen Lennard-Jones mixture~\cite{kob1995testing} in both two-dimensional (2D) and three-dimensional (3D) systems. The model consists of binary particle mixtures with a composition ratio of $80:20$ in 3D and $65:35$ in 2D, respectively. The interaction potential between particles is given by,

\begin{footnotesize}
\begin{equation}
\nonumber
V_{\alpha \beta}(r) = 4 \epsilon_{\alpha \beta}\left[\left(\frac{\sigma_{\alpha \beta}}{r}\right)^{12} - \left(\frac{\sigma_{\alpha \beta}}{r}\right)^{6} + \sum_{i=0}^{2} c_{2i}\left(\frac{r}{\sigma_{\alpha \beta}}\right)^{2i}\right].
\end{equation}
\end{footnotesize}
Here, $\alpha, \beta \in {A, B}$ denote the two types of particles, yielding three distinct interaction pairs. The potential is truncated at $r^{\alpha\beta}c = 2.5 \sigma_{\alpha \beta}$. The length scale is defined by $\sigma_{AA} = 1$, and the energy scale by $\epsilon_{AA} = 1$ (with Boltzmann's constant set to unity). The remaining parameters are $\epsilon_{AB} = 1.5$, $\epsilon_{BB} = 0.5$, $\sigma_{AB} = 0.8$, and $\sigma_{BB} = 0.88$.

To prepare solids at their shear minima, we use the LAMMPS package~\cite{thompson2022lammps}, employing a triclinic periodic simulation box. During energy minimization, macroscopic stress components are controlled using LAMMPS box/relax' command~\cite{lammps}. This protocol incorporates six (3D) or three (2D) box degrees of freedom in addition to interparticle potentials in the Hamiltonian, adjusting the box dimensions to achieve the desired diagonal stress values while tilting to control the off-diagonal stresses. Configurations at the minimum with respect to all shear degrees of freedom are obtained by constraining both off-diagonal ($\sigma_{\alpha\beta} = 0$) and diagonal ($\sigma_{\alpha\alpha} = 0$) stress components. 

The vibrational properties of the solids are investigated using the standard Hessian matrix, defined as 
\begin{equation}
    \left.\mathbf{H}_{i j} \equiv \frac{\partial^{2} V\left(\mathbf{r}_{1}, \mathbf{r}_{2}, \ldots \mathbf{r}_{n}\right)}{\partial \mathbf{r}_{i} \partial \mathbf{r}_{j}}\right|_{\left\{\mathbf{r}_{i}\right\}}, \label{eqn_hess}
\end{equation} where $V\left(\mathbf{r}_{1}, \mathbf{r}_{2}, \ldots, \mathbf{r}_{n}\right)$ is the total potential energy of the system, and $\mathbf{r}_{i}$ denotes the position of particle $i$. The eigenvectors of the Hessian matrix correspond to the cooperative displacements of the particles involved in the vibrations (referred to as normal modes), while the eigenvalues represent the vibrational energy, equal to $\omega^2$, where $\omega$ is the vibrational frequency. Eigenvalues are computed using the sparse solver \texttt{mkl\_sparse\_d\_ev} from the Intel Math Kernel Library~\cite{intelmkl}.

\section{Stability of Two-Dimensional Solids Under All Shear Deformations}\label{both_deformation}
\begin{figure}
\centering
\includegraphics[width=0.450\textwidth]{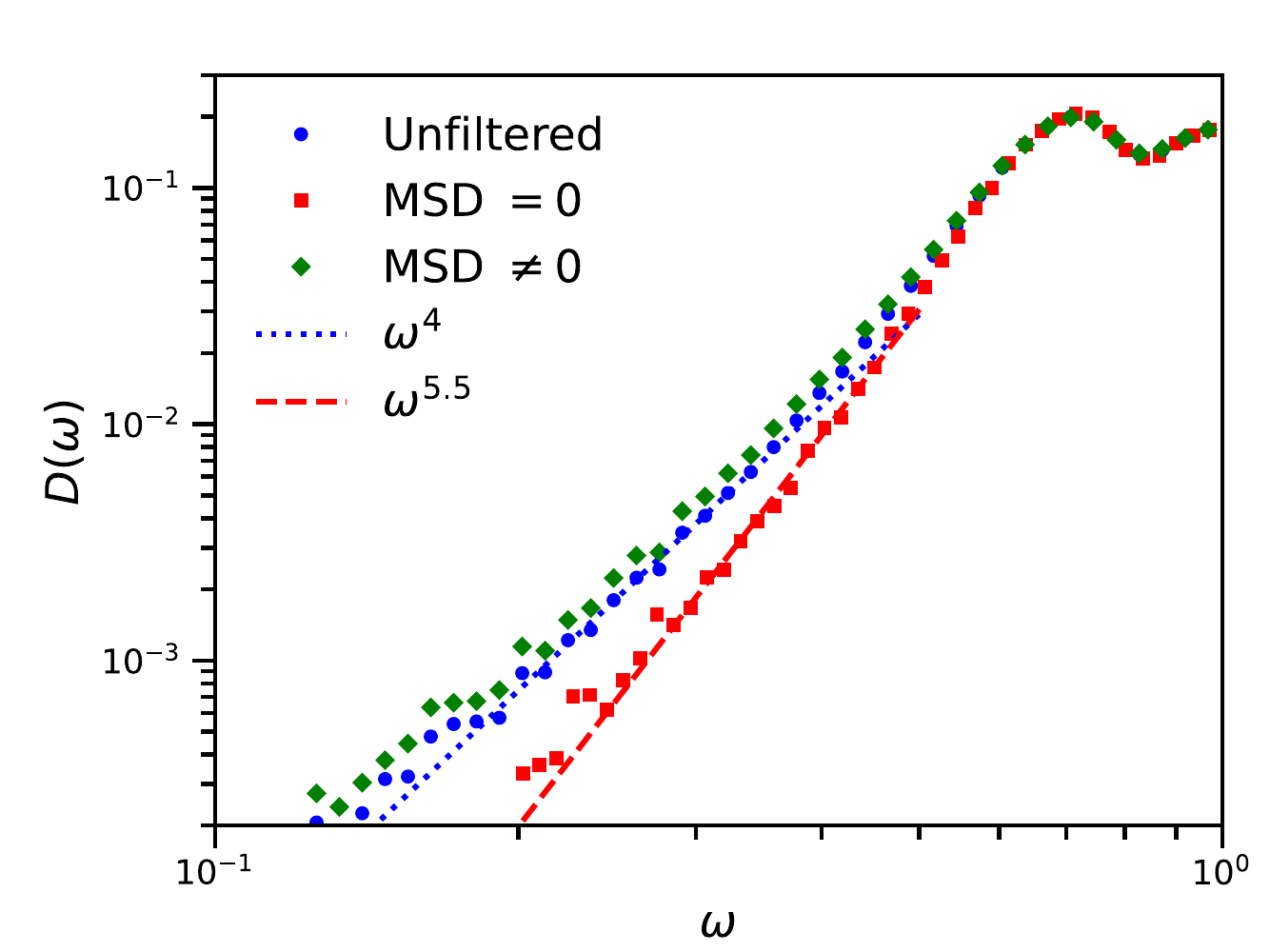}
\caption{
Low-frequency vibrational density of states (VDoS) for two-dimensional solids with system size $N=1024$ under simple and pure shear deformations. Configurations within the True elastic branch, where shear minima are accessible in all shear directions, exhibit a VDoS scaling behavior of $\omega^{5.5}$, indicative of a stable state. Conversely, configurations within the Fictitious elastic branch, where shear minima are inaccessible without plastic rearrangements, display a VDoS with a low-frequency power-law distribution, characterized by an exponent of approximately $4$.
}
\label{both_shear}
\end{figure}
We analyze the stability of two-dimensional solids under both simple and pure shear deformations. To achieve configurations where all shear stresses are constrained to zero, we iteratively rescale the box lengths and apply additional straining in the simple shear direction until fluctuations in all shear stress components fall below $10^{-8}$.

Configurations within the True elastic branch of both simple and pure shear are those where shear minima are accessible through deformation in all shear directions. In contrast, configurations that reside in the Fictitious elastic branch of any shear direction have inaccessible shear minima, requiring transitions to a lower-energy branch via plastic rearrangements. This distinction is determined by measuring the mean squared displacement (MSD) between the initial configurations and those obtained by performing AQS in the reverse direction, i.e., from $\sigma_{xy}=0$ to $\gamma_{xy}=0$, followed by rescaling the box dimensions to their original values.

Configurations with $\text{MSD}=0$ exhibit a VDoS scaling behavior of $\omega^{5.5}$, while those with a non-zero MSD display a low-frequency power-law distribution with an exponent of approximately $4$, as illustrated in Fig.~\ref{both_shear}.

\begin{figure}
\centering
\includegraphics[width=0.45\textwidth]{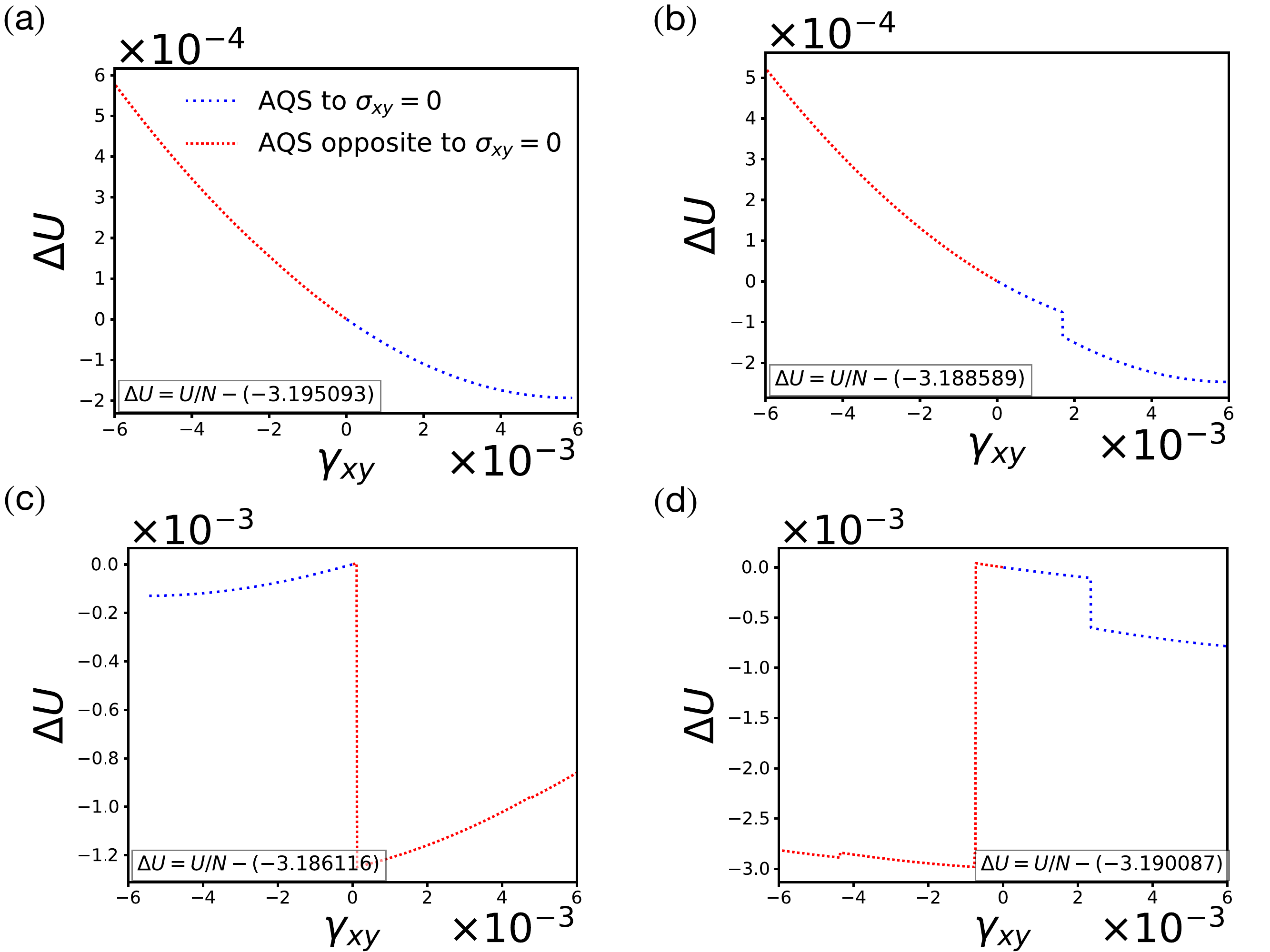}
\caption{
Energy versus strain response for a typical solid configuration with system size $N=1024$, generated using standard numerical protocols with periodic boundary conditions in two dimensions under the Athermal Quasistatic Shear (AQS) protocol. The blue line represents the response when straining toward the shear minima, while the red line shows straining in the opposite direction. \textbf{(a)} The configuration does not exhibit plastic deformation as $\delta \gamma \rightarrow 0$. EH analysis confirms it as stable, and the shear minimum is accessible through elastic deformation, classifying this configuration as $\mathcal{S}T$. \textbf{(b)} A configuration that remains stable (no plastic event) as $\delta \gamma \rightarrow 0$, as confirmed by an EH analysis. However, the shear minimum cannot be accessed through elastic deformation alone, requiring an irreversible plastic rearrangement to reach a lower energy branch. These configurations are EH-stable but reside on a Fictitious branch, classified as $\mathcal{S}F$. \textbf{(c)} The behavior of a configuration on a True simple shear elastic branch, which undergoes a plastic event when strained in the opposite direction from the shear minimum. This configuration is identified as unstable in the EH analysis. \textbf{(d)} A typical configuration on the Fictitious simple shear elastic branch, also classified as unstable in the EH analysis. 
}
\label{AQS_EH}
\end{figure}

\section{Extended Hessian}\label{appen_EH}
The elements of a standard Hessian matrix capture the local curvature of the energy landscape by considering only the particle positions as variables. Minima in this landscape correspond to stable, energy-minimized packings. However, when solids are prepared under periodic boundary conditions, the system may appear stable if only particle positions are considered, while it might actually be unstable when boundary deformations (such as shear strains, compressions, or expansions) are allowed~\cite{dagois2012soft}. These instabilities arise because the solid may not have fully relaxed with respect to macroscopic degrees of freedom. To properly assess the stability of a solid under periodic boundary conditions, an Extended Hessian (EH) matrix is introduced~\cite{dagois2012soft,goodrich2014jamming}. This EH matrix incorporates not only the particle positions but also the $d(d+1)/2$ degrees of freedom for boundary deformations. The EH matrix can be represented as,

\begin{equation}
    \mathbf{H}^{e}=\left[\begin{matrix}
\mathbf{H}^{e}_{rr} & \mathbf{H}^{e}_{r\epsilon} \\
\mathbf{H}^{e}_{\epsilon r} & \mathbf{H}^{e}_{\epsilon \epsilon}
\end{matrix}\right].
\end{equation}
where $\epsilon$ corresponds to different strain components. We can represent the individual elements of the matrix as,
\begin{equation}
    \begin{aligned}
        \mathbf{H}^{e}_{r_i^{a}, r_j^{b}}  &= \frac{\partial^2 V_{ij}}{\partial r^a_i \partial r^b_j},\\
        \mathbf{H}^{e}_{r_i^{a}, \epsilon_{bc}}  &= \sum_{j=1}^{N} \frac{\partial^2 V_{ij}}{\partial r_i^{a} \partial \epsilon_{bc}} ,\\
        \mathbf{H}^{e}_{\epsilon_{ab}, \epsilon_{cd}}  &= \frac{1}{2}\sum_{i=1}^{N}\sum_{j=1}^{N} \frac{\partial^2 V_{ij}}{\partial \epsilon_{ab} \partial \epsilon_{cd}}.
    \end{aligned}
\end{equation}

\begin{figure}
\centering
\includegraphics[width=0.49\textwidth]{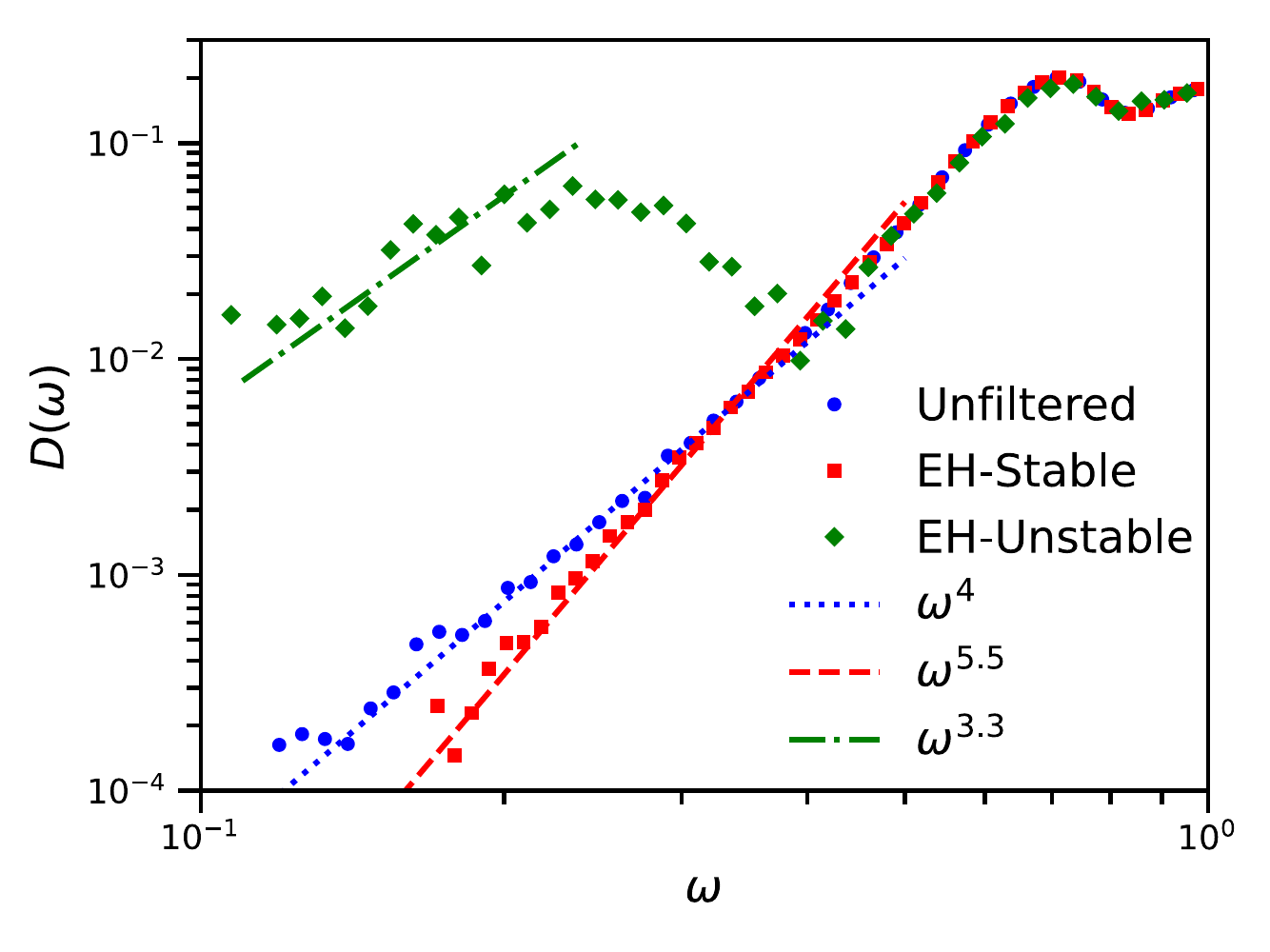}
\caption{(\textbf{a}) Low-frequency vibrational density of states (VDoS) for stable and unstable solids with system size $N=1024$ in two dimensions, as identified via an EH analysis. Stable solids exhibit a scaling behavior of $D(\omega) \sim \omega^{5.5}$, while unstable solids show $D(\omega) \sim \omega^{3.3}$. An ensemble comprising both stable and unstable solids presents a scaling behavior of $D(\omega) \sim \omega^{4}$.}
\label{VDoS_EH}
\end{figure}

In certain packings, the EH matrix may exhibit negative eigenvalues, indicating unstable configurations under infinitesimal boundary deformations. Solids on True elastic branches may undergo plastic instabilities when deformed away from their shear minima, potentially appearing unstable in EH analysis. Conversely, solids that appear stable under EH analysis can still reside on Fictitious branches, as EH identifies configurations prone to plastic instability only within a limited range of deformations. Fig.~\ref{AQS_EH} illustrates these different scenarios, showing the energy versus strain behavior for a typical solid configuration with system size $N=1024$, generated using standard numerical protocols with periodic boundary conditions in two dimensions under the AQS protocol. The red line represents the response when straining toward the shear minima, while the blue line shows straining in the opposite direction. \textbf{(a)} The configuration does not exhibit plastic deformation as $\delta \gamma \rightarrow 0$, and EH analysis confirms it as stable. The simple shear minimum is accessible through elastic deformation, classifying this configuration as $\mathcal{S}T$. \textbf{(b)} shows the behavior of a solid that does not undergo a plastic event as $\delta \gamma \rightarrow 0$, with EH analysis indicating stability. However, the shear minimum cannot be reached via elastic deformation alone, requiring irreversible plastic rearrangement to access a lower energy branch. These configurations are EH-stable but reside on a Fictitious branch, thus classified as $\mathcal{S}F$. \textbf{(c)} depicts the behavior of a configuration on the True simple shear elastic branch, which undergoes a plastic event when subjected to shear in the opposite direction from the shear minimum. This configuration is identified as unstable in the EH analysis. \textbf{(d)} shows a typical configuration on the Fictitious simple shear elastic branch, also classified as unstable in the EH analysis.

A recent study by Xu {\it et al.}~\cite{xu2024low} has demonstrated that EH-stable and unstable solids exhibit distinct power-law scaling behaviors in their low-frequency VDoS.  In Fig.\ref{VDoS_EH}, we present similar power-law scaling for solids classified as stable or unstable based on EH analysis, specifically for a system size of $N=1024$ in two dimensions. Stable solids exhibit a scaling behavior of $D(\omega) \sim \omega^{5.5}$, while unstable solids $D(\omega) \sim \omega^{3.3}$. Furthermore, the influence of instabilities from solids residing on Fictitious elastic branches-an aspect that EH analysis may not capture due to its restricted deformation domain-further modify the power-law behavior. EH-stable solids on the True elastic branch, under both simple and pure shear deformations, show a steeper power-law behavior with $D(\omega) \sim \omega^{6.5}$. In contrast, solids on Fictitious branches exhibit power-law scaling with an exponent greater than $5.5$. 

\end{appendix}


\bibliography{VDoS_Bibliography}

\end{document}